\documentclass[fleqn,10pt]{wlscirep}
\usepackage[utf8]{inputenc}
\usepackage[T1]{fontenc}
\usepackage{wrapfig}
\usepackage{caption}
\usepackage{multirow}

\title{Storyteller: The papers co-citing Sleeping Beauty and Prince before awakening}

\author[1,*]{Takahiro Miura}
\author[1]{Ichiro Sakata}
\affil[1]{School of Engineering, The University of Tokyo, Tokyo, Japan}
\affil[*]{miura@ipr-ctr.t.u-tokyo.ac.jp}


\begin{document}
\maketitle

\thispagestyle{empty}
\section*{Introduction}

In the Cumulative Advantage(CA) model, which is one of the most fundamental approaches to understand the mechanism of citation dynamics, papers receive citations depending on how much they have been already cited\citep{barabasi1999emergence}. On the other hand, a substantial effect not included in CA is that some surprising discoveries suddenly acquire citations after a long time from publishing. This phenomenon is known as Sleeping Beauty(SB)\citep{van2004sleeping}. Since disrupting discoveries need long-time discussion by the research community to accept\citep{barber1961resistance}, SBs can capture innovative findings and reveal the nature of disruptive scientific knowledge production. 
To research SB’s citation burst mechanism, bibliometricians consider the existence of the Prince(PR) for each SBs, which can be the trigger of SB’s awakeness. For example, the discovery of Green Fluorescent Protein(GFP), which got Nobel prize in chemistry, had been overlooked for 30 years until Chalfie and Tsien, who also received the prize, developed a method to use GFP as a marker protein in genetic engineering. However, how does Chalfie’s and Tsien’s research relight the hidden knowledge in the research community? If we can clarify such a mechanism rediscovering from nearly nothing, it can be helpful in science support and policy decision-making. This study proposes a “Storyteller” that focuses on the connection between SB and PR before SB gets citation burst by co-citation. PR is found to be the paper awakening SB in retrospect, but it is not easy to detect it as the trigger of SB’s awakeness at the time of PR submission. We named the papers which co-cites SB and PR before the citation burst of SB as “Storyteller”(ST) and analyze (1) how ST contributes to broadening the novelty of SB\&PR connections and (2) how much ST leads the citation burst after awakening.

\section*{Data and Methods}

This study adopted the 72 million papers metadata extracted from the Scopus Custom dataset (Jan.1970 - Oct.2020). To discuss SB’s leading to entirely discoveries rather than a summary of knowledge, the document types of SB’s were limited to “article” or “conference papers”, which were 62 million papers. First, we extracted high-impact SB’s according to Beauty Coefficient \citep{ke2015defining,miura2021large}. 
From the top 5\% of field and time corrected citations, we got 31,223 SB’s with the top 1\% Beauty Coefficient. Second, within the forward-citation of SB($N_{sb}$), the PR was defined as the most co-cited paper with the SB by 2020 published before the year of each SB’s awakening time $t_a$ calculated by \citep{ke2015defining}. In other words, PR is a paper that realized the importance of SB before it awakened and later played a vital role in the discovery of SB. Although a part of SB's had no prince as Zong mentioned\citep{zong2018sleeping}(133 SB’s which did not get any citations before their burst and 233 SB’s which did not have co-cited papers before their awakening), almost all SBs had corresponding PRs. The document types of PRs were mostly “article” (27,703), which means that SB has been found not as a summary of knowledge but as a combination of discovery. 
Lastly, the STs were defined as papers co-citing SB and PR before $t_a$. To check the amount of ST for each finding, we estimated the probability function of the number of ST for SB and PR with more than ten citations between $y_{pr}$ and $t_a$(23,243, 24,141 papers each).
To understand how STs contribute to future papers, we used the concept of disruptiveness \citep{wu2019large}. If papers citing ST after $t_a$ also cites SB more frequently than papers citing only SB-citing papers or PR-citing papers, STs incrementally contribute to the findings bring other research attention to SB. Given that SB and PR could get citations much later from citation burst, we limited the SB-citing and PR-citing papers to be compared to those submitted between the PR submission year $y_{pr}$ and $t_a$ same as ST.

\section*{Results}
\begin{figure}[h]
\begin{minipage}{0.55\textwidth}
\begin{center}
\includegraphics[width=\textwidth]{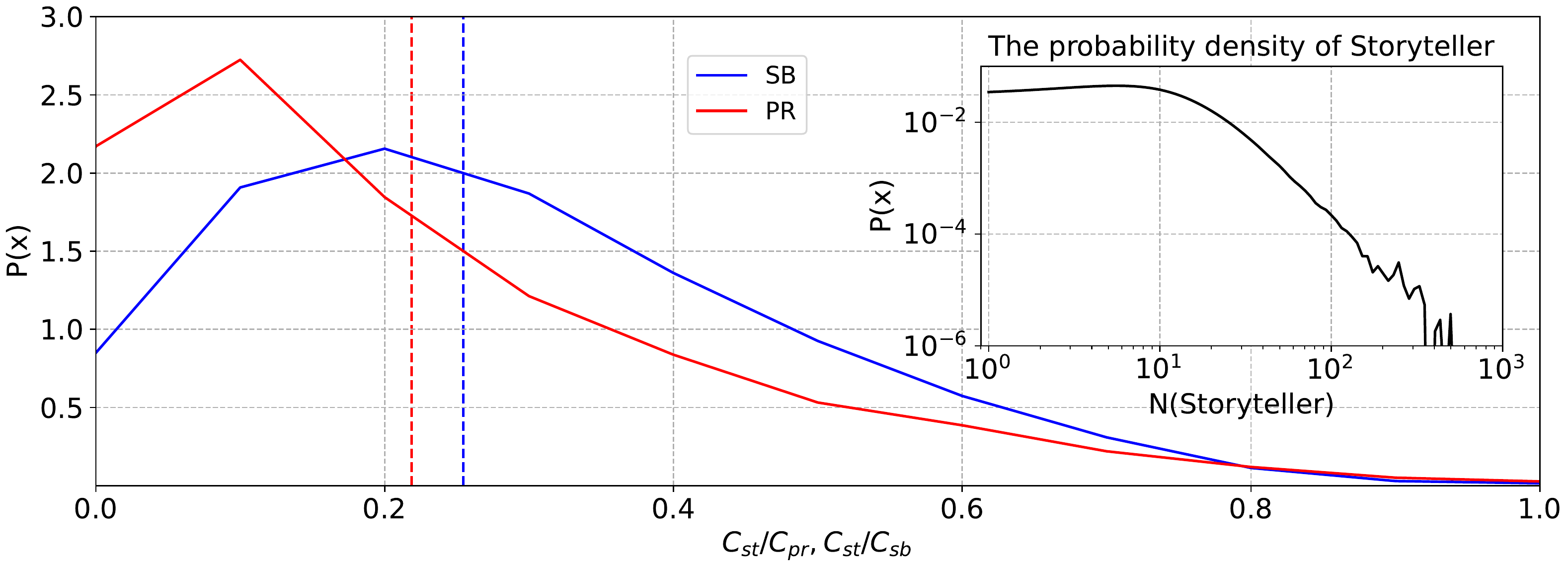}
\caption{The probablity density of the ratio of Storyteller in SB citation/PR citation at $t_a$($C^{sb}_{y_{pr}-t_a}>10$ or $C^{pr}_{y_{pr}-t_a}>10$) by Gaussian kernel density estimation, dotted line indicates the mean value. Inner graph shows the probablity density function of the number of Storyteller.}
\label{fig:num_of_st}
\end{center}
\end{minipage}
\hspace{20pt}
\begin{minipage}{0.4\textwidth}
\begin{center}
\makeatletter
\def\@captype{table}
\makeatother
\begingroup 
\renewcommand{\arraystretch}{1.4}
\begin{tabular}{cccc} \Hline
\multirow{2}{*}{Group($y_{pr}\leqq t \leqq t_a$)} & \multicolumn{3}{c}{Group($t > t_a$)} \\ \cline{2-4}
                                & $E_{SB}$ & $E_{PR}$ & $E_{|N_{sb}|}$\\ \hline 
$N_{sb} \cap \overline{N_{pr}}$ & 0.085 & 0.252 & 9.84\\
$\overline{N_{sb}} \cap N_{pr}$ & 0.047 & 0.072 & 22.02\\
$N_{sb} \cap N_{pr}$(=ST)       & \textbf{0.104} & 0.146 & 1\\ \Hline
\end{tabular} 
\endgroup
\caption{The mean probability of SB,PR citations after $t_a$ for papers satisfying ($C^{sb}_{y_{pr}-t_a}>10$) $\cap$ ($C^{pr}_{y_{pr}-t_a}>10)$. $E_{|N_{sb}|}$ is a relative number of papers citing SB after $t_a$ based on the number of papers citing both SB and ST.}
\label{tab:citation-type}
\end{center}
\end{minipage}
\end{figure}

Comparing the number of ST with the total number of SB citations between $y_{pr}$ and $t_a$($C^{sb}_{y_{pr} - t_a}$), 25.5\% of $C^{sb}_{y_{pr}-t_a}$ correspond to the ST on average(Figure.\ref{fig:num_of_st}). Similarly, ST occupies 21.9\% of $C^{pr}_{y_{pr}-t_a}$, indicating that identifying the link between SB\&PR before SB's citation burst is not an easy task.
Looking at the propagation rate of STs, STs are more likely to incrementally spread the findings of SB compared with papers citing only SB or PR(Table.\ref{tab:citation-type}). Moreover, they also propagate the PR papers than $\overline{N_{sb}} \cap N_{pr}$, meaning that STs could focus the research community's attention on a particular topic after the burst. Papers belonging to $N_{sb} \cap \overline{N_{pr}}$ also contribute to PR propagation incrementally.
On the other hand, the number of papers citing ST after $t_a$ is more than 9 or 22 times smaller than $N_{sb} \cap \overline{N_{pr}}$ and $\overline{N_{sb}} \cap N_{pr}$ because the number of STs are fewer, as Figure.\ref{fig:num_of_st} explains. In other words, although ST encourages future scientists to focus on the connection between SB and PR incrementally, disruptive SB and PR citations cause the SB citation burst.

\begin{figure}[h]
\vspace{-10pt}
    \centering
    \includegraphics[width=\textwidth]{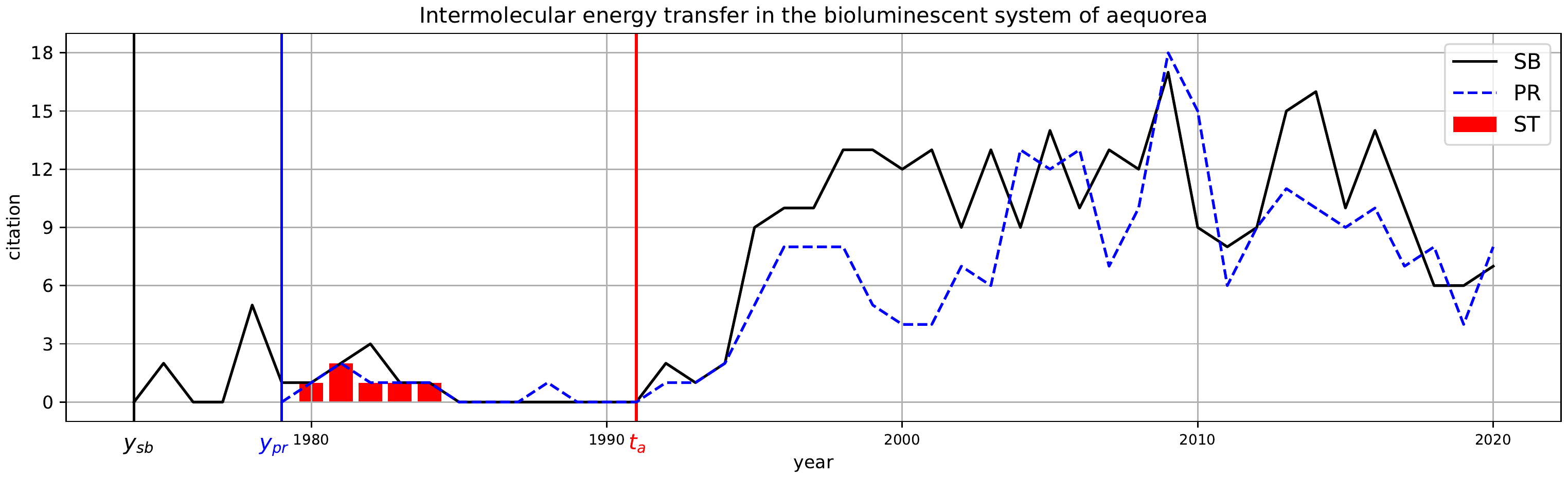}
    \caption{The citation history of Shimomura's SB and its PR. Red bar is the number of ST.}
    \label{fig:shimomura}
\vspace{-10pt}
\end{figure}

As a case study, take the example of Shimomura's Nobel-winning papers(Figure.\ref{fig:shimomura}). The method successfully extracted his papers "Intermolecular energy transfer in the bioluminescent system of aequorea(1974)" as SB; hence the PR was also Shimomura's "Structure of the chromophore of Aequorea green fluorescent protein(1979)", not Chalfie's nor Tsien's. The STs were published intensively immediately after PRs and accounted for all of the PR citations. As soon as $t_a$, SB began to rapidly increase its citations, suggesting that the trigger paper appeared immediately after $t_a$. 
Moreover, two of the six STs were SBs simultaneously, and Tsien's Nobel-winning papers defined by \citep{ioannidis2020work} was one of their STs. It means that SB and ST continue in alternating cycles while asleep, suggesting that a sudden significant discovery will bring both previous SB and ST to the public's attention at once.


\section*{Discussion and Conclusion}

Our study indicates that ST is only a small part of the papers that cite SB or PR before its discovery, but ST has more power to draw attention to SB and PR connections and make other researches focus on the critical connection between them than papers that only cite SB or PR.
However, the direct cause of the SB burst would not depend on ST; rather the innovative papers appear just around $t_a$.
Similar to phase transitions in physics, the state of the research evaluation before and after the awakening seems to have changed completely.
This research regard papers before $t_a$ as PR, but the key is at or just after the moment of $t_a$.
Future work will analyze who writes STs and how other PRs near $t_a$ lead to citation burst.
Through this research, we aim to make SB more of a scientific phenomenon rather than a terminology \citep{sugimoto2018note}.

\bibliography{main}

\end{document}